\documentclass{article}
\usepackage{spconf,amsmath,graphicx,amssymb}
\usepackage{multirow}
\usepackage{makecell}
\usepackage{tabularx}
\usepackage{booktabs}
\usepackage{hyperref}

\hypersetup{
    colorlinks=true,
    urlcolor=blue, 
    linkcolor=black,
    filecolor=black,      
    pdftitle={Your Document Title},
}

\usepackage{enumitem}
\setlist{nosep, leftmargin=14pt}

\usepackage{mwe} 

\title{T2-only prostate cancer prediction by meta-learning from bi-parametric MR imaging}
\name{
\parbox{\linewidth}{\centering
Weixi Yi$^{1,2}$,
Yipei Wang$^{1,2}$,
Natasha Thorley$^3$,
Alexander Ng$^4$,
Shonit Punwani$^3$,
Veeru Kasivisvanathan$^4$,\\
Dean C. Barratt$^{1,2}$,
Shaheer U. Saeed$^{1,2}$,
Yipeng Hu$^{1,2}$}}

\address{$^1$UCL Hawkes Institute, University College London, London, UK \\
$^2$Department of Medical Physics and Biomedical Engineering, University College London, London, UK \\
$^3$Centre for Medical Imaging, University College London, London, UK\\
$^4$Division of Surgery \& Interventional Science, University College London, London, UK}

\begin{document}
%
\maketitle

\begin{abstract}
Current imaging-based prostate cancer diagnosis requires both MR T2-weighted (T2w) and diffusion-weighted imaging (DWI) sequences, with additional sequences for potentially greater accuracy improvement. However, measuring diffusion patterns in DWI sequences can be time-consuming, prone to artifacts and sensitive to imaging parameters. While machine learning (ML) models have demonstrated radiologist-level accuracy in detecting prostate cancer from these two sequences, this study investigates the potential of ML-enabled methods using only the T2w sequence as input during inference time. We first discuss the technical feasibility of such a T2-only approach, and then propose a novel ML formulation, where DWI sequences - readily available for training purposes - are only used to train a meta-learning model, which subsequently only uses T2w sequences at inference. Using multiple datasets from more than 3,000 prostate cancer patients, we report superior or comparable performance in localising radiologist-identified prostate cancer using our proposed T2-only models, compared with alternative models using T2-only or both sequences as input. Real patient cases are presented and discussed to demonstrate, for the first time, the exclusively true-positive cases from models with different input sequences. Open-source code is available at \url{https://github.com/wxyi057/MetaT2}.
\end{abstract}
\begin{keywords}
Prostate cancer prediction, Bi-level optimisation, Meta learning, Diffusion model
\end{keywords}

\section{Introduction}
\label{sec:intro}
Various MR sequences are used in the image-based diagnosis of prostate cancer. Among these sequences, T2-weighted (T2w) images typically have shorter acquisition times, compared to other commonly used sequences such as diffusion-weighted imaging (DWI) with multiple b-values and dynamic contrast enhanced sequences. This fast acquisition is partly because efficient and robust T2w protocols are widely established and relatively standardised. For comparison, DWI protocols for prostate cancer diagnosis rely on specialised experience and, currently, exhibit higher variability in terms of sensitivity to prostate cancer, potentially because of variable factors including signal-to-noise ratio, susceptibility and rectal air artifacts \cite{giganti2020prostate}. The adoption of a T2-only diagnostic approach could substantially reduce the cost both in required expertise and imaging time, considerably streamlining prostate cancer patient care, compared with the recently-proposed machine learning (ML)-based approaches that require two or more sequences, e.g. \cite{lin2024evaluation, yan2024combiner}.

We argue that enabling T2-only prostate cancer diagnosis is not only desirable but also feasible, especially with modern ML models. First, DWI-visible tumours that are invisible (or challenging to identify by radiologists) on T2w were previously estimated to be $\sim$30\% of all clinically significant cases \cite{haider2007combined, lim2009prostate}. However, since no correlation has been established between radiologist false positives and DWI lesion invisibility (and such correlation is unlikely to be perfect), the actual number of cases missed by omitting DWI should be lower. This estimate is further complicated by the different yet unknown number of cases in radiologist false positive and false negative, which may be corrected by ML models (e.g. by learning underutilised inter-sequence and/or T2-to-cancer correlations). Furthermore, It is conceivable that the use of a single T2-only modality circumvents a number of challenges in multiparametric MR diagnosis, such as ground-truth label ambiguity due to inter-sequence spatial misalignment \cite{yang2022cross}, variable DWI protocols and quality discussed above. Therefore, it is worth investigating ML models to localise prostate cancer on T2w images, but maintaining competitive accuracy over models using multiple sequences.

This work focuses on developing a system that can best localize radiologist-identified cancer using only T2w images, in contrast to conventional diagnostic approaches that rely on multiple sequences. This is considered the first step to understand the feasibility of the T2-only approach, before developing models that are trained on and predict for histopathology labels. This step enables us to directly assess the model's capability to identify and localize cancerous regions from T2w images using radiologist annotations, without the additional complexity introduced by histopathological ground-truth sampling uncertainty \cite{bulten2020automated}. Such understanding is crucial for localization tasks requiring pixel-level annotations and may inform future development of histopathology-predicting models. Models developed in this work to predict radiologist labels can, on their own, enable a number of clinically useful applications including automating radiologist readings, and providing second opinions or assisting consensus during a radiology examination. 

To maximise the potential of predicting prostate cancer on T2w images, we propose 1) utilising existing DWI sequences during training - a form of learning using privileged information \cite{vapnik2009new}, 2) maximising the utilisation of the inference-omitted DWI sequences using conditional denoising diffusion models \cite{xia2024diffusion}, and 3) developing a new meta-learning framework to effectively train and predict prostate cancer using only T2w as input during inference. This concludes our contributions in \textit{a)} proposing a new clinical application, \textit{b)} developing a new technical ML algorithm, and \textit{c)} a set of rigorous evaluation experiments, with open-source code, using three cohorts from more than 3,000 prostate cancer patients, tested on both a publicly available PROMIS dataset \cite{ahmed2017diagnostic} and an internal multicentre MR-Targeted data set from multiple clinical trials \cite{hamid2019smarttarget, simmons2018accuracy, orczyk2021prostate, dickinson2013multi}.

\section{Methods}
\label{sec:method}
\begin{figure}[htbp]
   \centering
   \includegraphics[width=\linewidth]{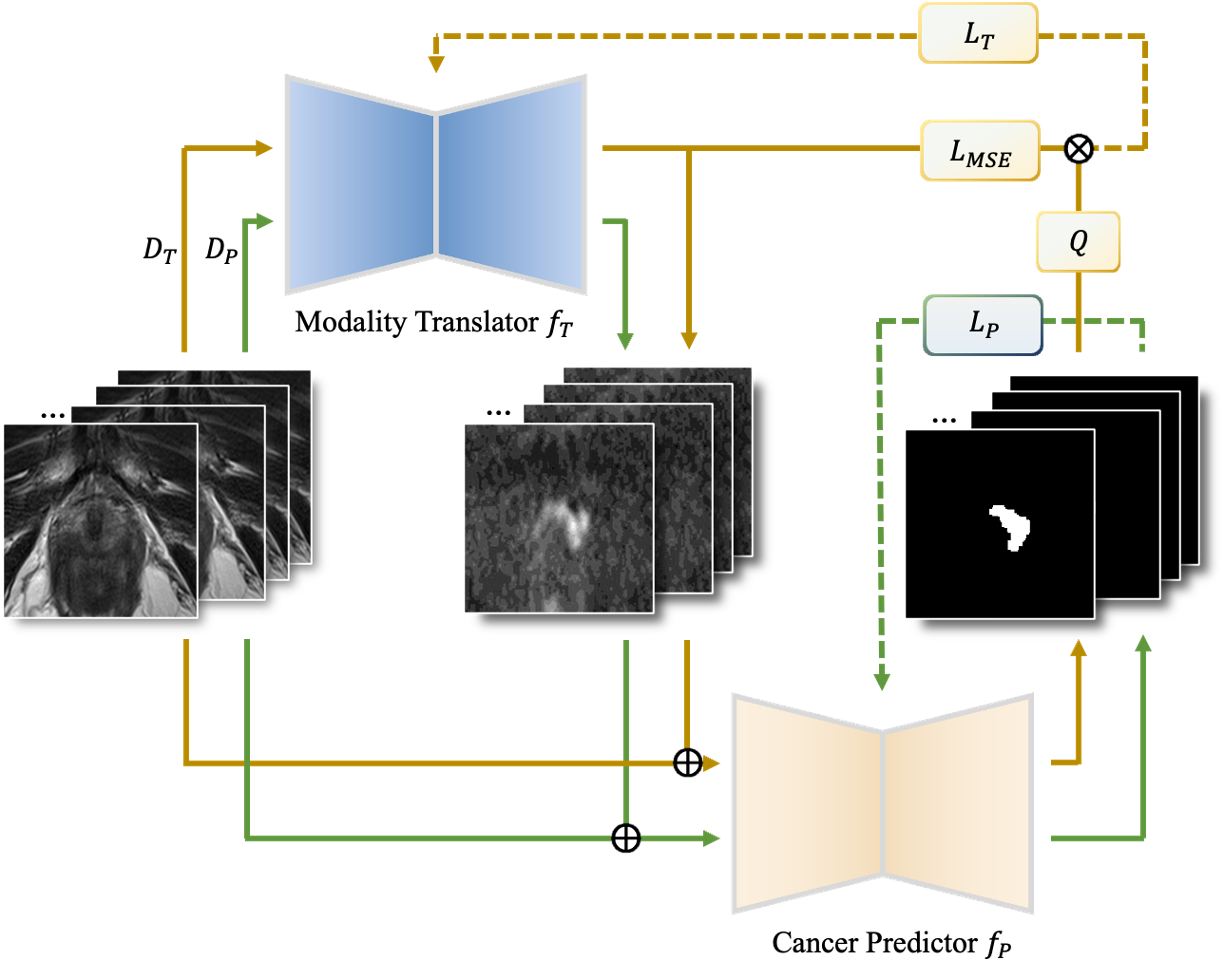}
   \caption{Illustration of the proposed bi-level meta learning framework MetaT2. Details are described in Sec.~\ref{sec:method}.}
   \label{fig:method}
\end{figure}

We propose a bi-level meta-learning framework that alternately optimizes two key components: a modality translator generating synthetic target modality $\hat{\mathcal{M}}_t$ from source modality $\mathcal{M}_s$, and a cancer predictor leveraging both $\mathcal{M}_s$ and $\hat{\mathcal{M}}_t$ for regions of interest (ROI) prediction. Figure~\ref{fig:method} illustrates our pipeline. The training dataset $D_{\text{train}} = {(\mathcal{M}_s, \mathcal{M}_t, \mathcal{G})}$ contains source modality images $\mathcal{M}_s$, target modality images $\mathcal{M}_t$, and ROI ground truth $\mathcal{G}$. We randomly split $D_{\text{train}}$ into equal-sized subsets $D_{\text{T}}$ and $D_{\text{P}}$ for training the modality translator and cancer predictor respectively.

\subsection{Modality translator: T2w-conditioned diffusion}
A modality translator $f_{\text{T}}(\cdot; \theta): \mathcal{M}_s \rightarrow \mathcal{M}_t$ learns a mapping from a source modality $\mathcal{M}_s$ to a target modality $\mathcal{M}_t$, where $\theta$ represents the learnable parameters. 
In this work, we use the diffusion model translator (DMT) based on the diffusion probabilistic diffusion model (DDPM) \cite{ho2020denoising} \cite{xia2024diffusion}. Once trained, the translator utilizes a diffusion process to generate a synthetic target modality $\hat{\mathcal{M}}t = f_{\text{T}}(\mathcal{M}_s; \hat{\theta})$ from the given source modality $\mathcal{M}_s$, where the target modality $\mathcal{M}_t$ in this work is DWI with high b-values. Training is performed using paired real images sampled from both $\mathcal{M}_s$ and $\mathcal{M}_t$, denoted as $\{m_s^k, m_t^k\}_{k=1}^N$, where $N$ represents the total number of sample pairs. The modality translator is trained to optimize the quality of the generated $\hat{m_t^k}$, using both the real target modality $m_t^k$ and a measure of cancer predictor performance, detailed in Sec.~\ref{sec:method-meta}.

\subsection{Cancer predictor for lesion segmentation}
The cancer localisation tasks are tested with a segmentation network, $f_{\text{P}}(\cdot; \omega): (\mathcal{M}_s, \hat{\mathcal{M}}_t) \rightarrow \mathcal{G}$, with learnable parameters $\omega$, which takes as input the source modality $\mathcal{M}_s$ and the synthetic target modality $\hat{\mathcal{M}}_t$ generated by the modality translator. In this work, we use UNet \cite{ronneberger2015u} as the backbone for the cancer predictor. The objective of the predictor is to predict ROI which in this work is prostate lesion, denoted as $\hat{\mathcal{G}} = f_{\text{P}}((\mathcal{M}_s, \hat{\mathcal{M}}_t); \hat{\omega})$. The model is trained with ground truth $\mathcal{G}$ sampled from real data, expecting to perform well on the synthetic target modality $\hat{\mathcal{M}}_t$ with good quality, further discussed in Sec.~\ref{sec:method-meta}.

\subsection{MetaT2: a bi-level meta-learning algorithm}
\label{sec:method-meta}
Our bi-level meta learning framework iteratively optimizes both the modality translator and the cancer predictor in a coordinated manner, enabling each network to benefit from the improvements of the other. This process involves two main phases, also known as inner and outer loops in bi-level optimization terminology: optimizing the cancer predictor and optimizing the modality translator.

\subsubsection{Optimizing the cancer predictor}

Let $L_{\text{Dice}}$ denote the Dice loss, which measures the overlap between the prediction and the ground truth $\mathcal{G}$, thus the cancer predictor is optimized by minimising loss $L_{\text{P}}$:

\[
L_{\text{P}} = \mathbb{E}_{(\mathcal{M}_s) \sim D_{\text{P}}}L_{\text{Dice}}\left(f_{\text{P}}((\mathcal{M}_s, f_{\text{T}}(\mathcal{M}_s; \theta)); \omega), \mathcal{G}\right)
\]

The network aims to improve the performance in predicting ROI with the synthetic modality $f_{\text{T}}(\mathcal{M}_s; \theta)$.

\subsubsection{Optimizing the modality translator}

The loss function $L_{\text{MSE}}$ measures the mean-squared-error (MSE) between the output of the modality translator $f_{\text{T}}(\mathcal{M}_s; \theta)$ and the target modality $\mathcal{M}_t$. The modality translator is optimised using a joint loss that combines the $L_{\text{MSE}}$ and a quality score $Q$. $Q=L_{\text{Dice}}\left(f_{\text{P}}((\mathcal{M}_s, f_{\text{T}}(\mathcal{M}_s; \theta)); \omega), \mathcal{G}\right)$, is computed with respect to the cancer predictor, representing the accuracy of the prediction on the synthetic modality. The overall loss function thus is:
\[
L_{\text{T}} = \mathbb{E}_{(\mathcal{M}_s, \mathcal{M}_t) \sim D_{\text{T}}} \left[ \alpha L_{\text{MSE}} \left(f_{\text{T}}(\mathcal{M}_s; \theta), \mathcal{M}_t \right) + \left(1-\alpha\right) Q \right]
\]
where $\alpha$ is the hyperparameter balancing between the $L_{\text{MSE}}$ and the quality score $Q$, experimented in Sec.3.3.

\subsubsection{Bi-level optimisation of the networks}

To jointly train the cancer predictor $f_{\text{P}}$ and the modality translator $f_{\text{T}}$, we employ a bi-level optimization framework that consists of two nested optimization loops: an inner loop and an outer loop. The inner loop optimizes the modality translator's parameters $\theta$ to generate synthetic modalities that aims to improve the cancer predictor performance. The outer loop optimizes the cancer predictor's parameters $\omega$
, aiming for accurate cancer prediction while considering the improvements provided by the modality translator. The optimization problem is formulated as:
\[
\begin{aligned}
&\omega^* = \arg\min_{\omega} L_{\text{P}}(\omega, \theta^*(\omega)),\\
& \text{s.t.} \, \theta^*(\omega) = \arg\min_{\theta} L_{\text{T}}(\omega, \theta)
\end{aligned}
\]

Here, $\theta^*(\omega)$ represents the optimal parameters of the modality translator obtained from the inner loop for a given set of cancer predictor parameters $\omega$. Training alternates between updating $\theta$ to minimize translator loss $L_{\text{T}}$ given $\omega$ in the inner loop, and updating $\omega$ to minimize the predictor loss $L_P$ using the updated $\theta^*(\omega)$ in the outer loop.

\section{Experiments}

\begin{table*}
\centering
\caption{Comparisons of different methods for prostate MR lesion segmentation on MR-Targeted and PROMIS datasets.}
\scriptsize
\begin{tabularx}{\textwidth}{@{}c*{2}{c}*{4}{>{\centering\arraybackslash}X}*{4}{>{\centering\arraybackslash}X}@{}}
\toprule
\multirow{2}{*}{Method} & \multirow{2}{*}{Train} & \multirow{2}{*}{Test} & \multicolumn{4}{c}{MR-Targeted} & \multicolumn{4}{c}{PROMIS} \\
\cmidrule(lr){4-7} \cmidrule(l){8-11}
& & & DSC$\uparrow$ & HD95$\downarrow$ & Rec.$\ast^{\text{GT}}$
$\uparrow$ & Prec.$\ast^{\text{Pd}}$$\uparrow$ & DSC$\uparrow$ & HD95$\downarrow$ & Rec.$\ast^{\text{GT}}$ $\uparrow$ & Prec.$\ast^{\text{Pd}}$$\uparrow$ \\
\midrule
UNet & T2w, DWI & T2w, DWI & 0.3322(0.2679)&23.30(22.63) & 0.3980&0.3116 &0.3156(0.2612) &24.56(23.29) & 0.3305 & 0.2391 \\
\midrule
UNet & T2w & T2w &0.2844(0.2172) & 27.59(12.02)& 0.5011&0.1324 & 0.3124(0.2228)&26.22(12.20) & 0.4555 & 0.1279 \\
nnUNet & T2w & T2w &0.3035(0.2840) & 29.51(31.49)&0.3025 &\textbf{0.3594} &0.3209(0.2882) &26.76(28.23) & 0.2968 & \textbf{0.2947} \\

DDPM+UNet & T2w, DWI & T2w & 0.3038(0.2685)&25.92(25.81) &0.3453 &0.2693 &0.2973(0.2639) &24.98(23.45) & 0.3198 & 0.2161 \\
Ours(MetaT2) & T2w, DWI & T2w &\textbf{0.3392(0.2358)} &\textbf{21.46(11.76)} &\textbf{0.5397} &0.2862 & \textbf{0.3351(0.2359)}&\textbf{22.48(11.44)}  & \textbf{0.4786} & 0.2240 \\
\bottomrule
\label{tab:comparison}
\end{tabularx}
\end{table*}

\subsection{Dataset}

\textbf{MR-Targeted:} This dataset includes multiparametric T2w and DWI MR images from 850 patients across UK clinical trials led by UCLH \cite{hamid2019smarttarget, simmons2018accuracy, orczyk2021prostate, dickinson2013multi}. Among these, 689 patients with PIRADS$\geq$3 lesions were manually annotated by radiologists and split into 275/275/139 for training the modality translator, cancer predictor, and testing. All 3D images were resampled to $0.3 \text{mm}\times 0.3 \text{mm} \times 1 \text{mm}$ resolution, and lesion-containing slices were extracted and center-cropped to $256 \times 256$ pixels. Details are in \cite{yan2024combiner}.

\textbf{PROMIS:} The PROMIS dataset is a publicly available multicenter study comprising multiparametric 3D MR images from 566 patients blind to prior biopsy~\cite{ahmed2017diagnostic}. Lesions were contoured by radiologists on reports and annotated on T2w images by a biomedical imaging researcher. Following the same preprocessing as MR-Targeted dataset, we used this dataset exclusively to test our proposed method.

\textbf{PICAI:} This dataset contains 1285 multiparametric 3D prostate MR images from patients who underwent MR and ultrasound-guided biopsy\cite{saha2023artificial}. Following the same preprocessing as MR-Targeted dataset, we used DWI from this dataset to pre-train DDPM in the modality translator.

\subsection{Evaluation Metrics}

We evaluated performance using both voxel and lesion-level metrics. At voxel level, we used Dice similarity coefficient (DSC) for overlap and 95$^{th}$ percentile Hausdorff Distance (HD95) for boundary accuracy. At lesion level, following \cite{yan2024combiner}, we calculated: (1) Lesion-level Recall Rec.$\ast^{\text{GT}}$, where a ground-truth lesion is counted as true positive if sufficiently overlapping with any prediction; (2) Lesion-level Precision Prec.$\ast^{\text{Pd}}$, where a predicted lesion is counted as true positive if sufficiently overlapping with any ground truth.

\subsection{Implementation Details}
The DDPM was pre-trained on PICAI following Nichol et al.\cite{nichol2021improved}'s ImageNet settings. For cancer prediction, UNet and nnUNet \cite{isensee2021nnu} served as benchmarks. The modality translator and cancer predictor were pre-trained on $D_T$ and $D_P$ for 60 and 180 epochs respectively, followed by 10 epochs of bi-level optimization. We used Adam optimizer with learning rate $1 \times 10^{-4}$ and set $\alpha=0.75$ for translator optimization. All experiments ran on NVIDIA Quadro GV100 GPUs.

\subsection{Comparison and ablation studies}

We evaluate the performance of our proposed MetaT2 framework by comparing it with the model trained under different modality configurations: (1) UNet (T2w, DWI): Trained and tested on both T2w and DWI images. (2) UNet (T2w): Trained and tested solely on T2w images. (3) nnUNet (T2w): Trained and tested solely on T2w images. (4) DDPM+UNet: A modality translator is first trained using paired T2w and DWI images to generate synthetic DWI. Then, a UNet is trained and tested using T2w images and the synthetic DWI.

\section{Results}

\begin{figure}[htbp]
   \centering
   \includegraphics[width=0.78\linewidth]{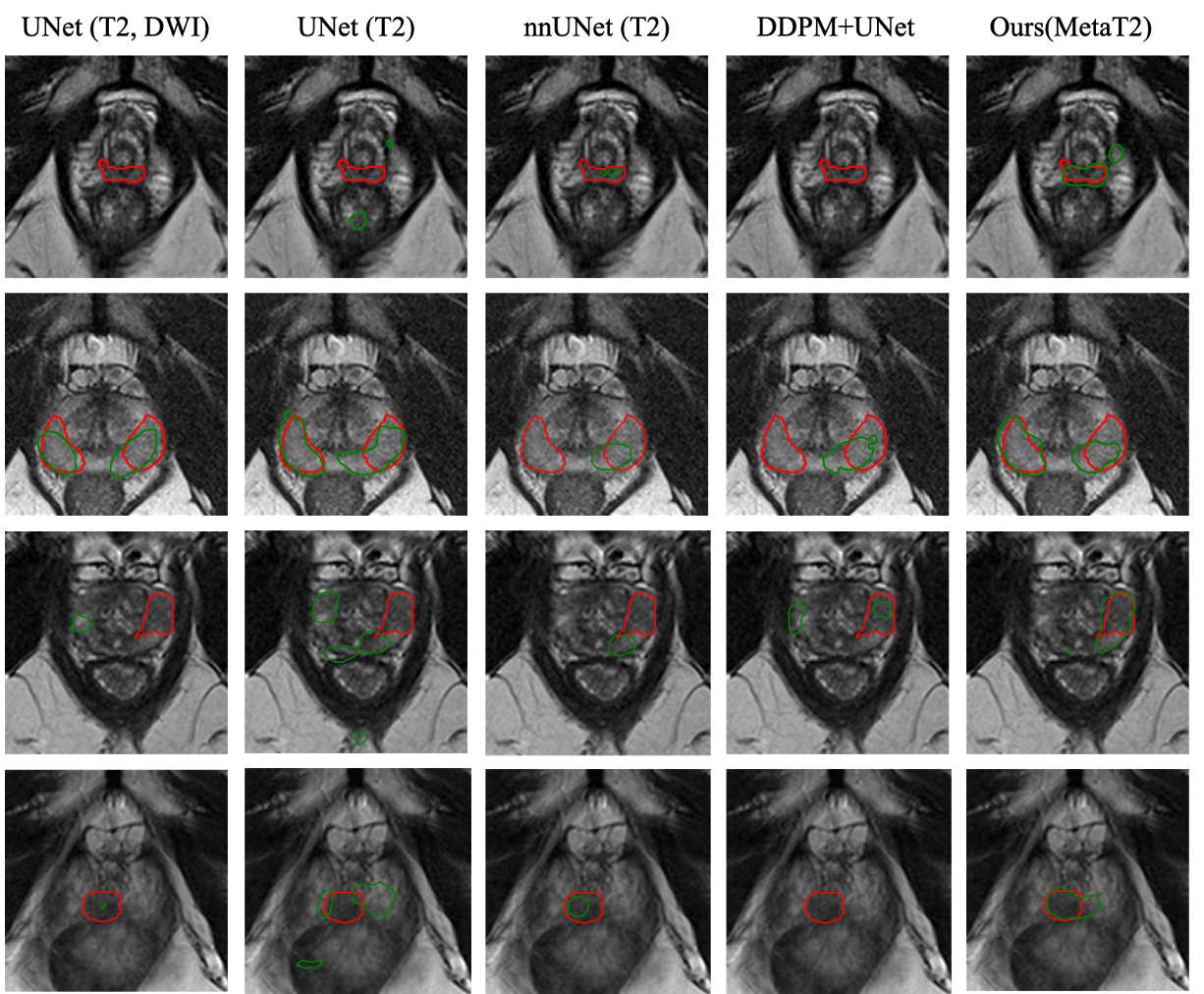}
   \caption{Samples with segmentation from the tested methods. The red and green contours represent ground truth and prediction, respectively.}
   \label{fig:results}
\end{figure}

Figure \ref{fig:results} shows example segmentation results from different methods. Table \ref{tab:comparison} summarizes quantitative results for prostate lesion segmentation on the MR-Targeted and PROMIS. Our proposed MetaT2 framework achieves the highest performance on both datasets, with a larger improvement on the unseen PROMIS dataset. UNet and nnUNet models trained only on T2w images obtain worse DSC and HD95 scores on MR-Targeted compared to UNet trained with both T2w and DWI images. This suggests that relying solely on T2w images limits segmentation accuracy, perhaps due to the lack of complementary information from DWI. However, MetaT2 achieves segmentation performance comparable to, or even surpassing, the UNet (T2w, DWI) method, despite requiring only T2w images at inference. Specifically, MetaT2 improves DSC by 2.1\% on the MR-Targeted dataset and 6.2\% on the PROMIS dataset compared to UNet (T2w, DWI). This may indicate that the proposed meta-learning effectively leverages multi-modal data during training to enhance segmentation performance in single-modality inference. The improvement may result from reduced false positives due to omitting DWI or other factors discussed in Sec.~\ref{sec:intro}. Consequently, this suggests that a diagnostic approach using only T2-weighted images is feasible without compromising accuracy.

The comparison between MetaT2 and DDPM+UNet serves as an ablation study on the effectiveness of the bi-level meta-learning framework. As shown in Table~\ref{tab:comparison}, \begin{figure}[htbp]
   \centering
   \includegraphics[width=0.62\linewidth]{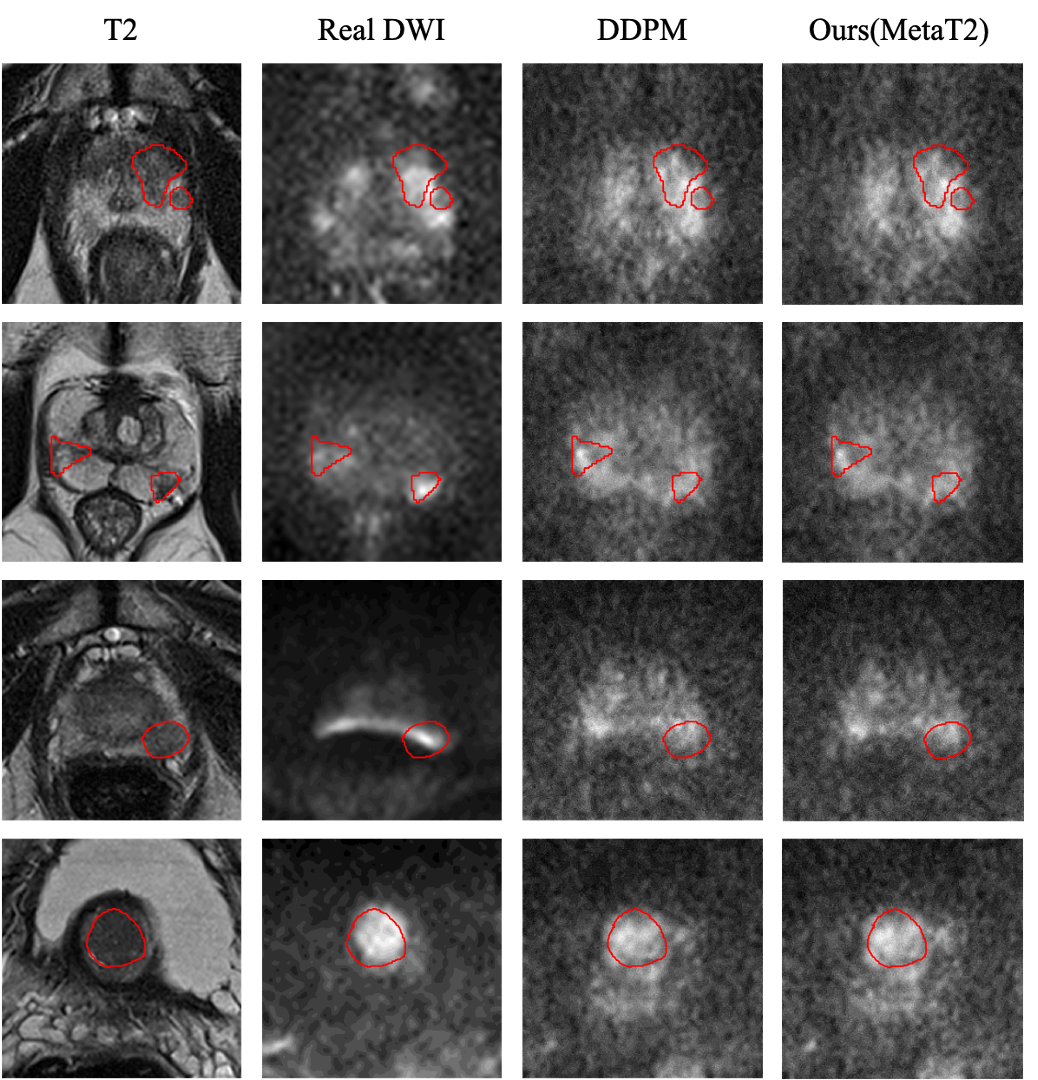}
   \caption{Samples of T2w images, corresponding real DWI, and DWI images generated by different methods. The red contours indicate the lesion ground truth.}
   \label{fig:syn_results}
\end{figure}incorporating bi-level meta-learning strategy improves the model performance. Compared to DDPM+UNet, which does not use the meta-learning framework, MetaT2 improves DSC for prostate lesion segmentation by 11.65$\%$ on the MR-Targeted dataset and 12.71$\%$ on the PROMIS dataset. 

MetaT2 provides the highest Rec.$\ast^{\text{GT}}$, achieving 0.5397 on MR-Targeted and 0.4786 on PROMIS datasets, surpassing all other methods. However, its lower precision Prec.$\ast^{\text{Pd}}$ is expected to indicate a trade-off between the improved lesion detection coverage and false positive predictions.

Figure \ref{fig:syn_results} shows the results of our modality translator. The synthesized DWI demonstrates comparable lesion signal characteristics and contrast to real DWI, while reducing common artifacts and distortions.

\section{Conclusion}


We have presented MetaT2, a bi-level meta-learning framework that enables accurate prostate cancer lesion prediction using only T2-weighted MRI data at inference. By leveraging DWI information during training and jointly optimizing the modality translator and the cancer predictor, MetaT2 achieves superior performance compared to tested baseline methods. Our results suggest that a T2-only diagnostic approach is both feasible and beneficial, potentially improving efficiency and accessibility in prostate cancer prediction.

\section{Compliance with Ethical Standards}
This study complies with Declaration of Helsinki. Ethics approval and patient consent were obtained for MR-Targeted dataset \cite{hamid2019smarttarget, simmons2018accuracy, orczyk2021prostate, dickinson2013multi}, and publicly available datasets (PROMIS \cite{ahmed2017diagnostic}, PICAI \cite{saha2023artificial}) were used per their terms.

\section{Acknowledgments}
This work is supported by the International Alliance for Cancer Early Detection, an alliance between Cancer Research UK [C28070/A30912; C73666/A31378], Canary Center at Stanford University, the University of Cambridge, OHSU Knight Cancer Institute, University College London and the University of Manchester.

\bibliographystyle{IEEEbib}
\bibliography{strings,refs}

\begin{thebibliography}{10}

\bibitem{giganti2020prostate}
Francesco Giganti, Clare Allen, et~al.,
\newblock ``Prostate imaging quality (pi-qual): a new quality control scoring system for multiparametric magnetic resonance imaging of the prostate from the precision trial,''
\newblock {\em European urology oncology}, vol. 3, no. 5, pp. 615--619, 2020.

\bibitem{lin2024evaluation}
Yue Lin, Enis~C Yilmaz, Mason~J Belue, et~al.,
\newblock ``Evaluation of a cascaded deep learning--based algorithm for prostate lesion detection at biparametric mri,''
\newblock {\em Radiology}, vol. 311, no. 2, pp. e230750, 2024.

\bibitem{yan2024combiner}
Wen Yan, Bernard Chiu, Ziyi Shen, et~al.,
\newblock ``Combiner and hypercombiner networks: Rules to combine multimodality mr images for prostate cancer localisation,''
\newblock {\em Medical Image Analysis}, vol. 91, pp. 103030, 2024.

\bibitem{haider2007combined}
Masoom~A Haider, Theodorus~H Van Der~Kwast, et~al.,
\newblock ``Combined t2-weighted and diffusion-weighted mri for localization of prostate cancer,''
\newblock {\em American journal of roentgenology}, vol. 189, no. 2, pp. 323--328, 2007.

\bibitem{lim2009prostate}
Hyun~Kyung Lim, Jeong~Kon Kim, et~al.,
\newblock ``Prostate cancer: apparent diffusion coefficient map with t2-weighted images for detection—a multireader study,''
\newblock {\em Radiology}, vol. 250, no. 1, pp. 145--151, 2009.

\bibitem{yang2022cross}
Qianye Yang, David Atkinson, Yunguan Fu, et~al.,
\newblock ``Cross-modality image registration using a training-time privileged third modality,''
\newblock {\em IEEE Transactions on Medical Imaging}, vol. 41, no. 11, pp. 3421--3431, 2022.

\bibitem{bulten2020automated}
Wouter Bulten, Hans Pinckaers, et~al.,
\newblock ``Automated deep-learning system for gleason grading of prostate cancer using biopsies: a diagnostic study,''
\newblock {\em The Lancet Oncology}, vol. 21, no. 2, pp. 233--241, 2020.

\bibitem{vapnik2009new}
Vladimir Vapnik and Akshay Vashist,
\newblock ``A new learning paradigm: Learning using privileged information,''
\newblock {\em Neural networks}, vol. 22, no. 5-6, pp. 544--557, 2009.

\bibitem{xia2024diffusion}
Mengfei Xia, Yu~Zhou, Ran Yi, et~al.,
\newblock ``A diffusion model translator for efficient image-to-image translation,''
\newblock {\em IEEE TPAMI}, 2024.

\bibitem{ahmed2017diagnostic}
Hashim~U Ahmed, Ahmed El-Shater Bosaily, Louise~C Brown, et~al.,
\newblock ``Diagnostic accuracy of multi-parametric mri and trus biopsy in prostate cancer (promis): a paired validating confirmatory study,''
\newblock {\em The Lancet}, vol. 389, no. 10071, pp. 815--822, 2017.

\bibitem{hamid2019smarttarget}
Sami Hamid, Ian~A Donaldson, Yipeng Hu, et~al.,
\newblock ``The smarttarget biopsy trial: a prospective, within-person randomised, blinded trial comparing the accuracy of visual-registration and magnetic resonance imaging/ultrasound image-fusion targeted biopsies for prostate cancer risk stratification,''
\newblock {\em European urology}, vol. 75, no. 5, pp. 733--740, 2019.

\bibitem{simmons2018accuracy}
Lucy~AM Simmons, Abi Kanthabalan, Manit Arya, et~al.,
\newblock ``Accuracy of transperineal targeted prostate biopsies, visual estimation and image fusion in men needing repeat biopsy in the picture trial,''
\newblock {\em The Journal of urology}, vol. 200, no. 6, pp. 1227--1234, 2018.

\bibitem{orczyk2021prostate}
Clement Orczyk, Dean Barratt, et~al.,
\newblock ``Prostate radiofrequency focal ablation (proraft) trial: a prospective development study evaluating a bipolar radiofrequency device to treat prostate cancer,''
\newblock {\em The Journal of Urology}, vol. 205, no. 4, pp. 1090--1099, 2021.

\bibitem{dickinson2013multi}
Louise Dickinson, Hashim~U Ahmed, et~al.,
\newblock ``A multi-centre prospective development study evaluating focal therapy using high intensity focused ultrasound for localised prostate cancer: the index study,''
\newblock {\em Contemporary clinical trials}, vol. 36, no. 1, pp. 68--80, 2013.

\bibitem{ho2020denoising}
Jonathan Ho, Ajay Jain, and Pieter Abbeel,
\newblock ``Denoising diffusion probabilistic models,''
\newblock {\em NeurIPS}, vol. 33, pp. 6840--6851, 2020.

\bibitem{ronneberger2015u}
Olaf Ronneberger, Philipp Fischer, et~al.,
\newblock ``U-net: Convolutional networks for biomedical image segmentation,''
\newblock in {\em MICCAI 2015}. Springer, 2015, pp. 234--241.

\bibitem{saha2023artificial}
Anindo Saha, Joeran Bosma, Jasper Twilt, et~al.,
\newblock ``Artificial intelligence and radiologists at prostate cancer detection in mri—the pi-cai challenge,''
\newblock in {\em MIDL}, 2023.

\bibitem{nichol2021improved}
Alexander~Quinn Nichol and Prafulla Dhariwal,
\newblock ``Improved denoising diffusion probabilistic models,''
\newblock in {\em ICML}. PMLR, 2021, pp. 8162--8171.

\bibitem{isensee2021nnu}
Fabian Isensee, Paul~F Jaeger, Simon~AA Kohl, et~al.,
\newblock ``nnu-net: a self-configuring method for deep learning-based biomedical image segmentation,''
\newblock {\em Nature methods}, vol. 18, no. 2, pp. 203--211, 2021.

\end{thebibliography}

\end{document}